\documentclass[pre]{revtex4}

\usepackage{epsfig}
\usepackage{epsf}
\usepackage{graphicx}
\usepackage{dcolumn}
\usepackage{bm}

\def\be{\begin{equation}}
\def\ee{\end{equation}}
\def\ba{\begin{array}}
\def\ea{\end{array}}
\def\bea{\begin{eqnarray}}
\def\eea{\end{eqnarray}}

\def\sign{{\mathrm{sign}}}

\begin{document}


\title{Laplacian transfer across a rough interface: \\
Numerical resolution in the conformal plane}

\author{Damien Vandembroucq and St\'{e}phane Roux}
 \email{damien.vandembroucq@saint-gobain.com; stephane.roux@saint-gobain.com}
\affiliation{Surface du Verre et Interfaces,\\
Unit\'e Mixte de Recherche CNRS/Saint-Gobain,\\
39 Quai Lucien Lefranc, 93303 Aubervilliers cedex, France.}


\begin{abstract}
We use a conformal mapping technique to study the Laplacian transfer
across a rough interface. Natural Dirichlet or Von Neumann boundary
condition are simply read by the conformal map. Mixed boundary
condition, albeit being more complex can be efficiently treated in the
conformal plane. We show in particular that an expansion of the
potential on a basis of evanescent waves in the conformal plane allows
to write a well-conditioned 1D linear system. These general principle
are illustrated by numerical results on rough interfaces.
\end{abstract}



\maketitle
\vspace*{-0.5cm}
\section{Introduction}

Various physical phenomena such as the diffusion of oxygen molecules
through lungs, the complex impedance of a rough electrode, the
heterogeneous catalysis on a rough or porous substrate can be
described by the simple model of Laplacian transport across irregular
boundaries\cite{Grebenkov-PhD04}. Although very simple in the case of
plane or smooth interfaces, the resolution of the problem becomes
extremely arduous as soon as the geometry of the boundary is
irregular. The interest for this question was stimulated when it
appeared that the irregularity of numerous surfaces could be
mathematically modelled as fractal and in the last two decades a large
number of works have been devoted to this question, notably by Sapoval
and coworkers, who clarified these different problems through seminal
contributions.

The fractal description of rough or porous electrodes allowed them to
obtain exact results in the case of deterministic fractal
electrodes\cite{Sapoval88PRA} and more generally this motivated the
use of scaling arguments \cite{Sapoval91PRA,Sapoval92PRA} in the study
of the constant phase angle (CPA) behavior of rough electrodes. More
recently, they showed that the frequential behavior of a rough
electrode could be obtained by studying the spectral properties of the
so-called self-transport
operator\cite{SapovalEPJB99a,SapovalEPJB99b}. The latter measures the
probability for two given sites of an interface to be linked by random
walk through the electrolyte. One of the difficulty raised by the
Laplacian transport through irregular interfaces comes from the nature
of a physically sound boundary condition (b.c.). Neither Dirichlet nor
Neumann type do apply, but rather a mixed b.c. holds quite generally,
namely $V=\Lambda \partial_n V$ where $V$ is the potentiel and
\overrightarrow{n} the outward normal to the interface. The length
$\Lambda$ is the ratio of the surface resistance of the electrode to
the resistivity of the electrolyte. To overcome this difficulty
Sapoval \cite{Sapoval94PRL} proposed to replace the mixed b.c. by a
Dirichlet condition on an equivalent interface obtained by coarse
graining at the scale $\Lambda$ from the original interface.

In the same spirit one can also think of replacing the b.c.  on the
rough interface by a derived one on a plane interface. In two
dimensions, this can be easily performed by a conformal
map\cite{VR-PRE97a,VR-PRE97b}. For a simple b.c. such as Dirichlet
(constant potentiel) or Von Neumann (constant flux), the harmonic
problem is entirely solved once the conformal map has been determined:
the solutions are precisely given by the real and imaginary parts of
the map from the rough electrode to the plane one. However, in case of
a mixed b.c., the situation becomes more complex: the ``simple'' mixed
b.c. on a rough interface has to be replaced by a ``heterogeneous''
b.c. on the equivalent plane interface. The ``heterogeneity'' of the
b.c. simply reflects the harmonic measure on the original interface. A
related difficulty arises when studying a Stokes flow along a rough
boundary\cite{VR-PRE97b}, the conformal map thus transforms the
original bi-Laplacian equation into a more complex equation where the
derivative of the conformal map acts as a non constant coefficient. In
the last two cases, however, one can show that the use of a conformal
map helps computing the potential in spite of the complexity of the
equivalent equation or boundary condition. Expanding the potential on
a basis of evanescent waves in the conformal plane allows us to write
a well conditioned linear system, a much more laborious work in the
original geometry. Let us emphasize that the conditioning (and hence
precision) of the formulation is the most salient interest of the
approach.

In the following, we recall how to compute a conformal map onto a
two-dimensional domain bounded by a single valued interface; then we
write down the equivalent linear system corresponding to the
resolution of the Laplacian transfer across a rough interface. We
finally present numerical results.

\section{Conformal map}

The conformal half-plane and the physical domain are referenced by the
complex coordinates $z=x+iy$ and $w=u+iv$ respectively and defined by
$y<0$ and $v<h(u)$, where $h$ is a known function periodic of period
$2\pi$. There exists a map $w=\Omega(z)$ such that the image of the
axis $y=0$, is the rough interface.  Equivalently, we can write
$\Im[\Omega^{-1}(u+ih(u))]=0$.

The map $\Omega$ is chosen under the form:
$\displaystyle \Omega(z)=z+2\sum_{k=0}^N \overline{a_k} e^{-i kz}$,
coordinate by coordinate, we have:

  \be\left\{\ba{lll}
  u&\displaystyle
  =x
  +\sum_{k=0}^N \overline{a_k} e^{-ik x}
  +\sum_{k=0}^N           a_k  e^{ ik x}
&\displaystyle=x+\sum_{k=-N}^N c_k  e^{ ik x}
\\
\label{eq:v}
  h(u)&\displaystyle=
   \sum_{k=0}^N \overline{(ia_k)} e^{-ik x}
  +\sum_{k=0}^N           (ia_k)  e^{ ik x}
&\displaystyle=\sum_{k=-N}^N b_k e^{ik x}
  \ea\right.\ee
with $b_k=ia_k$ for $k>0$,
$b_k=-i\overline{a_{-k}}=\overline{b_{-k}}$ for $k<0$, 
and $b_0=i(a_0-\overline{a_0})$. Simultaneously
 $c_k=a_k$ with $k>0$, $c_k=\overline{a_{-k}}$ for $k<0$, and 
$c_0=a_0+\overline{a_0}$.
Thus $c_k=-i\sign(k) b_k$

The computation of $a_k$ is made through the following steps: let us
assume as a starting hypothesis $x\approx u$. Then from
Eq.~\ref{eq:v}, $b_k$ is given by the Fourier transform of $h(x)$.  We
construct the $c_k$ as the Hilbert transform of $h$, and thus we have
access to a first corrected estimate of $u(x)$. Therefore for an
arithmetic sequence of $x$, an {\it unevenly} distribution sequence of
$u$ results, which is used to sample $h(u(x))$.  The iteration of this
step can be shown to converge as soon as the maximum local slope of
the interface is lower than unity\cite{VR-PRE97a}, and the latter
constraint can be relaxed using using under-relaxation at the expense
of computational efficiency.

 \begin{figure}[b]
 \centerline{\epsfig{file=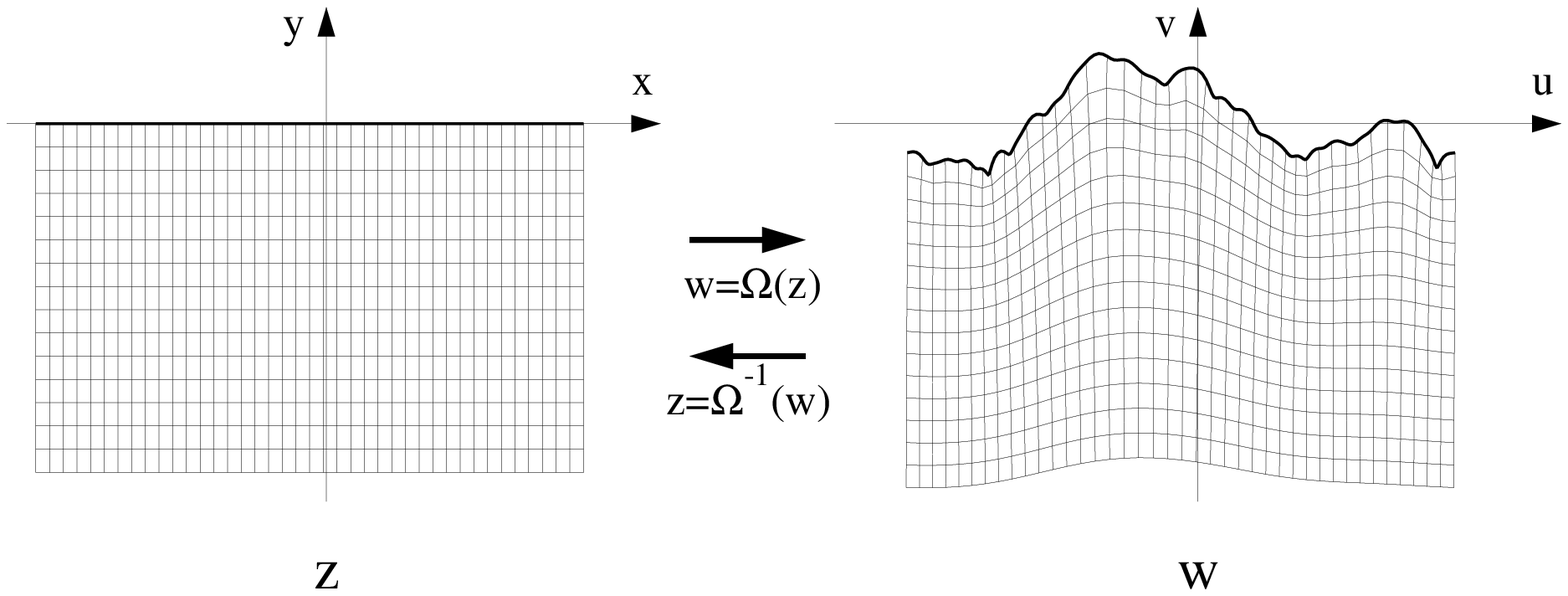,width=.8\hsize}}
 \caption{\label{fig1} Conformal map from the lower half-plane onto a
 domain bounded by an arbitrary (single valued) rough interface. The
 image of the regular mesh gives a direct access to the equipotential
 and current lines in the rough geometry.  }
 \end{figure}

 \begin{figure}[t]
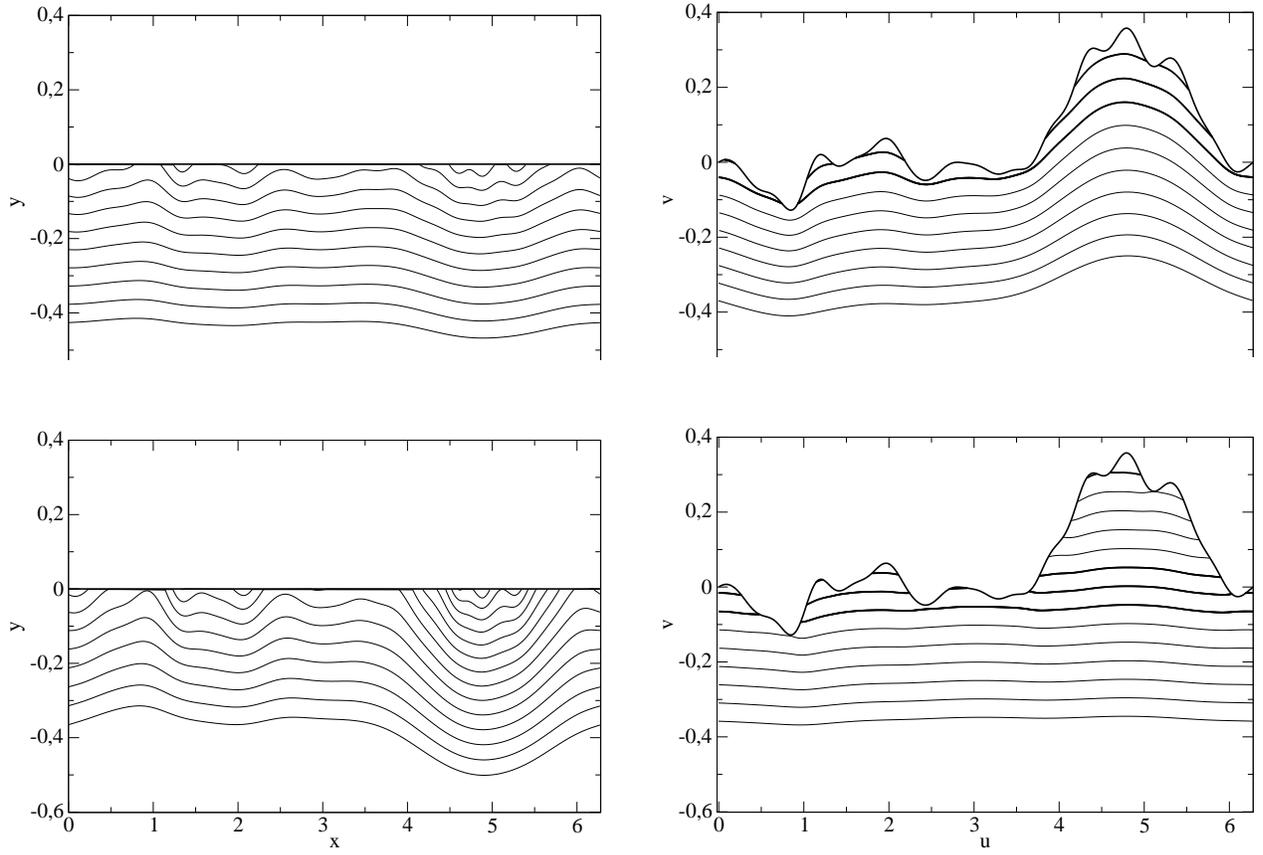

\centerline{\hfill
\epsfig{file=isoz025.eps,width=.44\hsize}
\hfill
\epsfig{file=isow025.eps,width=.44\hsize}\hfill
}
\centerline{\hfill
\epsfig{file=isoz4.eps,width=.44\hsize}
\hfill
\epsfig{file=isow4.eps,width=.44\hsize}\hfill
}
 \caption{\label{iso} Equipotential lines in the conformal geometry (left)
 and in the physical domain (right) for two mixed boundary conditions:
 $\Lambda=0.25$ (above) and $\Lambda=4$ (below). The bold curve
 represents the interface. }
 \end{figure}

\section{Use of the map for harmonic problem solutions}

The determination of the conformal map gives a direct solution for the
equipotential condition $V=V_0$ along the boundary and a prescribed
gradient $\partial_yV=\alpha$ far from the interface: $V=V_0+\alpha
y=V_0+\alpha \Im \left[\Omega^{-1}(w)\right]q$. Similarly, the
solution for zero flux flowing out of the boundary is obtained from
the real part of the inverse conformal map.  Other b.c. require some
more work.  In the case of a prescribed inhomogeneous potential,
Dirichlet condition, the field in the entire domain is obtained from a
single 1D Fourier transform of the imposed potentiel. The extension of
the solution to the bulk is naturally provided by using evanescent
modes in the conformal domain.

In the case of inhomegeneous Neumann b.c., one should take into
account the fact that in the mapping the gradients are transformed.
Let us first introduce the following notation concerning the gradient
of a scalar real function $A$.  The gradient is a vector of
coordinates $(\partial_u A, ~\partial_v A)$ , can be represented as a
complex number $\partial_u A + i\partial_v A=\partial_{\overline w}
A$.  In the transformed domain, we have
  \be
  \partial_{\overline z} A
  = \overline{\Omega'(z)} \partial_{\overline w} A
  \ee
For a prescribed flux, once the conformal map has been obtained,
the transformed flux in the $z$ plane is obtained from the above
formula.  Again looking for a decomposition of the potential over
a basis of exponential functions gives a straightforward answer.
For the case of mixed boundary conditions, the problem appears to
be somewhat more complicated.  We will consider a simple case of a
boundary condition written as
  \be
  V=\Lambda (\vec \nabla V)\vec n
  \ee

where $\vec n$ is the unit normal to the boundary, and where
$\Lambda$ is a characteristic length scale. In the far field,
$y\to-\infty$, we impose $\vec \nabla V\to \vec e_y$. In the
reference plane $z$, the normal component of the gradient is along
the $y$ axis, since the conformal map preserves angles. The
boundary condition written in the reference plane is thus
  \be
  |\Omega'(z)|V(z)=\Lambda (\partial_y V)
\label{BC}
  \ee
We search $V(z)$ as the real part of a sum of evanescent modes:
  \be
  V(z)=y+\alpha_0 +\Re\left[\sum_{n\ge 1} 2\alpha_n e^{-in z}\right]
  \ee

Because the conformal map has been determined in a first step, the
$|\Omega'(z)|$ is known.  We naturally introduce its Fourier
transform, so that
  \be
  |\Omega'(x,0)|=\beta_0 +\Re\left[\sum_{n\ge 1} 2\beta_n e^{-in
  x}\right]=
\beta_0 + \sum_{n=1}^N \beta_n e^{-in x} 
+  \sum_{n=1}^N \overline{\beta_n} e^{in x} \;.
  \ee

The satisfaction of the boundary condition (\ref{BC}) thus leads to:

\be
\alpha_0\beta_0+\sum_{k=1}^N \left(\alpha_k \overline{\beta_k}
+\overline{\alpha_k}\beta_k\right)=\Lambda 
\quad{\rm and}\quad
\left(A_{nk}-2\pi n \delta_{nk}\right) \alpha_k= 0\;
(n\ge1)
\ee

where $A_{nk}=\beta_{n-k}$ if $k\le n$ and $A_{nk}=\overline{\beta_{k-n}}$ 
if $k> n$.

\section{Numerical results}

\begin{figure}[]
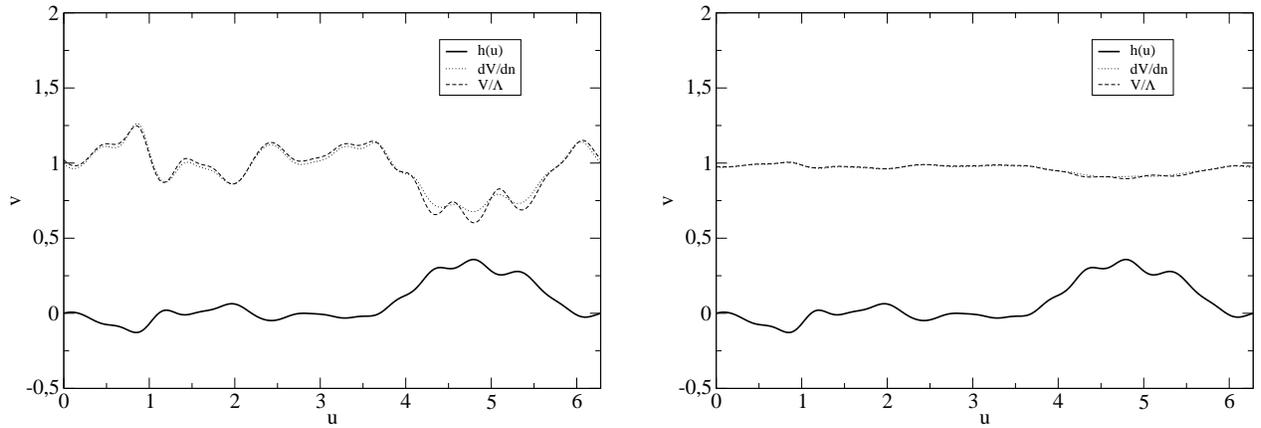

\centerline{\hfill
\epsfig{file=testCL.eps,width=.44\hsize}
\hfill
\epsfig{file=testCL4.eps,width=.44\hsize}\hfill
}
 \caption{\label{test} Satisfaction of the mixed boundary condition
 along the rough interface for the two mixed boundary conditions
 $\Lambda=0.25$ (left) and $\Lambda=4$ (right).}
 \end{figure}

We present here results obtained for a mixed boundary condition on a
rough interface characterized by a self-affine scaling of roughness
exponent $\zeta=0.8$ (with a lower cut-off =2$\pi$/16) and of maximum
local slope $s_{max}=0.75$ (Note that we only consider here an
intermediate self-affine scaling: the interface remains regular at
short scales). The Laplacian potential is written on a basis of 512
evanescent modes. Figure \ref{iso} shows the equipotential lines
obtained in the conformal half-plane $(x,y)$ and their images in the
physical domain $(u,v)$ for two different b.c. with $\Lambda=0.25$ and
$\Lambda=4$ respectively.  In Fig. \ref{test} we represent the normal
derivative of the potential and the surface potential sealed by the
length $\Lambda$. The satisfaction of the b.c. corrresponds to the
equality of these two quantities. We obtain very nice results except
for the lowest values of the potential for $\Lambda=0.25$. This is
attributed to the truncation of the high frequency modes performed in
writing the linear system. Let us note that the equipotential lines
shown in Fig. \ref{iso} define precisely the effective Dirichlet
b.c. giving rise to the same behavior (including near field). This
object is the one for which B. Sapoval \cite{Sapoval94PRL} proposed a
geometrical construction based on coarse graining.

\end{document}